\documentclass[aps,prl,superscriptaddress,preprint]{revtex4}
\usepackage{graphicx}
\usepackage[colorlinks=true,linkcolor=blue,citecolor=blue,urlcolor=blue]{hyperref}
\usepackage{xcolor}
\usepackage{amsmath}
\usepackage[capitalize]{cleveref}
\def\beq{\begin{equation}}
\def\eeq{\end{equation}}
\def\bea{\begin{eqnarray}}
\def\eea{\end{eqnarray}}

\begin{document}
\makeatletter
\title{Sounds of Leidenfrost drops} 

\author{Tanu Singla}
\email{tanu.singla@uaem.mx}
\affiliation{Centro de Investigaci\'on en Ciencias-(IICBA), UAEM, Avenida Universidad 1001, Colonia Chamilpa, Cuernavaca, Morelos, M\'{e}xico} 
\affiliation{Tecnol\'ogico de Monterrey, Calle del Puente 222, Colonia Ejidos de Huipulco, Tlalpan, Ciudad de M\'exico, M\'{e}xico}

\author{M. Rivera}
\affiliation{Centro de Investigaci\'on en Ciencias-(IICBA), UAEM, Avenida Universidad 1001, Colonia Chamilpa, Cuernavaca, Morelos, M\'{e}xico} 

\date{\today}

\begin{abstract}
We show that when a drop of water is maintained in its Leidenfrost regime, a sound in the form of periodic beats emits from the drop. The process of beat emission involves two distinct frequencies. One component is the frequency of beats itself and second is the frequency of sound in every beat which is emitted when one oscillation in the drop occurs. Experiments have been performed by placing a drop of water over a concave metallic surface and the beats of the drop were recorded by fixing a microphone above the drop. A video camera was also fixed above the drop to record its oscillations. Simple analytical techniques like Fourier and wavelet transforms of the audio signals and image processing of the videos of the drop have been used to gain insight about mechanism of beat emission process. This analysis also helped us in studying the dependence of frequencies, if any, on the radius of the drop and the substrate temperature.
\end{abstract}

\maketitle

Leidenfrost effect explains how a drop of liquid after coming in contact with a surface, at a temperature much higher than the boiling point of the liquid, floats over a cushion of liquid vapors between the drop and the surface. This effect is known to the scientific community since 1751 and it gained popularity in 1952 when Holter et al. \cite{Holter} accidentally found the oscillating patterns in the drop. Since then, interest in people to study this effect has not subsided and a plethora of research is being reported even until today; inverse Leidenfrost effect \cite{Hall}, Leidenfrost wheels \cite{Bouillant}, about fate of Leidenfrost droplets \cite{Lyu}, Leidenfrost levitation \cite{Hashmi}, granular Leidenfrost effect \cite{Eshuis},  Leidenfrost effect in hydrogels \cite{Waitukaitis,Waitukaitis1}, self propelled Leidenfrost drops \cite{Linke} are to name a few. 

Oscillations in liquid drops due to distinct perturbations in itself is a widely explored topic. It is known that under small perturbations, frequencies of the resonance modes of liquid drops follows the relation $f_d\propto r_d^{-3/2}$, where, $r_d$ is the radius of the drop \cite{Rayleigh}. Following this idea, Shen et al. studied the dynamics of acoustically levitated drops as a function of $r_d$ \cite{Shen}. This was followed by Bouwhuis et al. \cite{Bouwhuis} where they studied frequencies of oscillations of a liquid drop floating over an air cushion. In a completely different scenario, in 1996, Yoshiyasu et al. \cite{Yoshiyasu} studied the behavior of water drop on a vertically oscillating plate. One of the principle findings of this work was that the frequency of the oscillations of the drop was half of the frequency of perturbation i.e. the drop behaved like a parametric oscillator. Our group has also reported oscillations of liquid metal subjected to different perturbations \cite{Dinesh,Elizeth,Jorge,Singla}. Despite the fact that the liquid drops in these works were subjected to different perturbations, due to the identical underlying hydrodynamic principles, similar star shaped oscillations were observed in all cases.

In the case of Leidenfrost drops, the vapor layer trapped between the drop and the hot substrate plays a crucial role in its dynamics. The geometry and dynamics of the vapor layer have been studied using interference imaging technique \cite{Burton,Caswell} and the dependence of the vapor layer thickness on the drop radius was established. In another interesting study, the thickness of the vapor layer was calculated using electrostatic methods \cite{Carmes}. Ma et al. \cite{Ma} have reported different oscillating modes (``Leidenfrost Stars'') in various liquids. Contrary to popular belief, they further reported that in the case of Leidenfrost effect, the resonant frequencies of the drop do not depend on its radius. Moreover, similar to the drops on vertically vibrated plate, mentioned earlier, the frequency of the vapor layer beneath the drop has been reported to be double the frequency of oscillations of the drop.

Here we introduce a new phenomenon; the emission of sound from a Leidenfrost drop. We show that when a drop of water was kept on a substrate with temperature much higher than the boiling point of water, a periodic sound was emitted from the drop. From here onwards, we will mention periodic sound as beats. It was observed that, when the size of the drop was large, beats were emitted only when the drop oscillated in one of the star configuration modes. In contrast, beats were also detected in small drops even though the stars cannot be observed in drops of these sizes. It was further observed that the beats emitted from the drop (both large and small) can be decomposed into two frequencies: the frequency of the beats ($f_b$) which is related to the fix interval of time between two consecutive beats and the frequency of sound ($f_s$) emitted in every beat. In the present work, we study both components of frequencies ($f_b$ and $f_s$) as a function of the drop radius ($r_d$) and the substrate temperature ($T_s$).

\begin{figure*}[ht!]
\includegraphics[width=16cm,height=6cm]{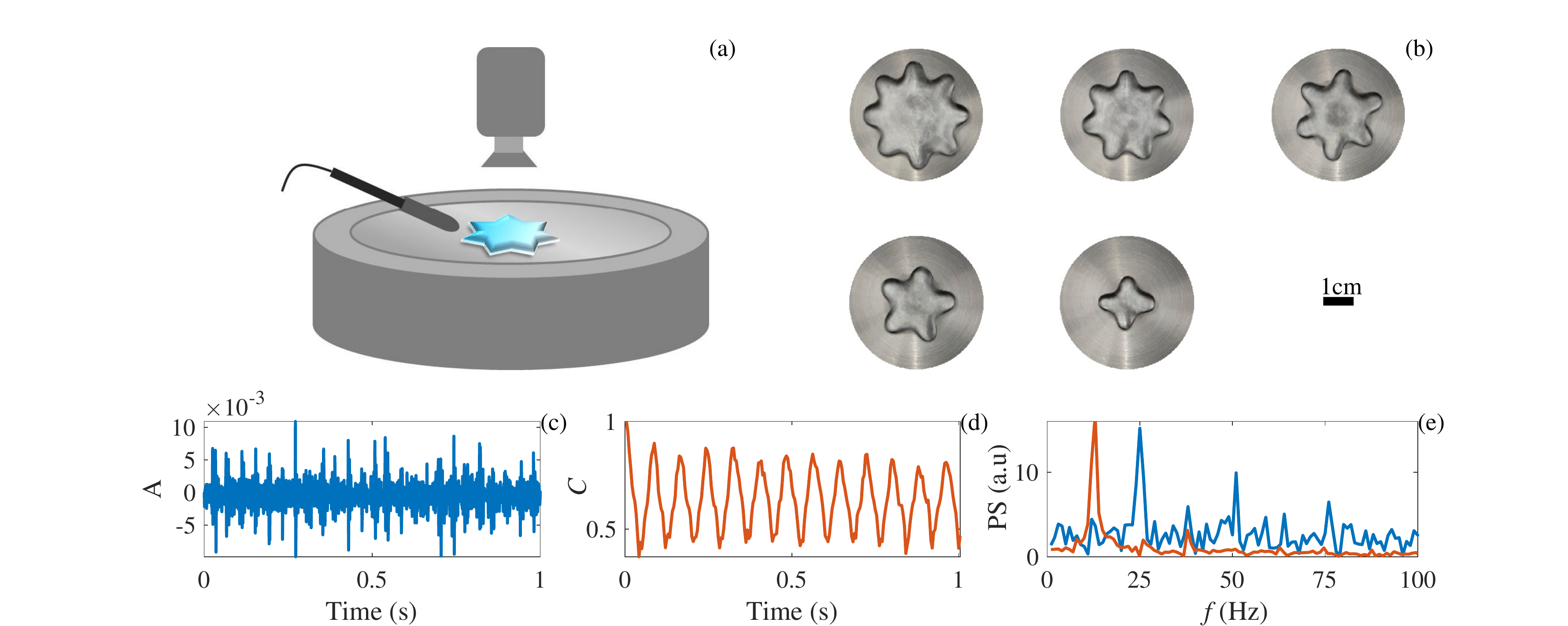}
\caption{(a) Experimental setup that was used to study the beats of Leidenfrost drops, (b) geometric patterns in the drop that result in the beat emission phenomenon, (c) an example of the audio signal of the beats emerging from the drop ($r_d\approx1.1$ cm), (d) time series corresponding to geometric oscillations in the drop, and (e) Fourier transforms of the time series shown in (c, and d). Temperature of the substrate ($T_s$) for these results was $500^\circ$ C.} 
\label{expsp}
\end{figure*}

\begin{figure}[ht!]
\includegraphics[width=8cm,height=6cm]{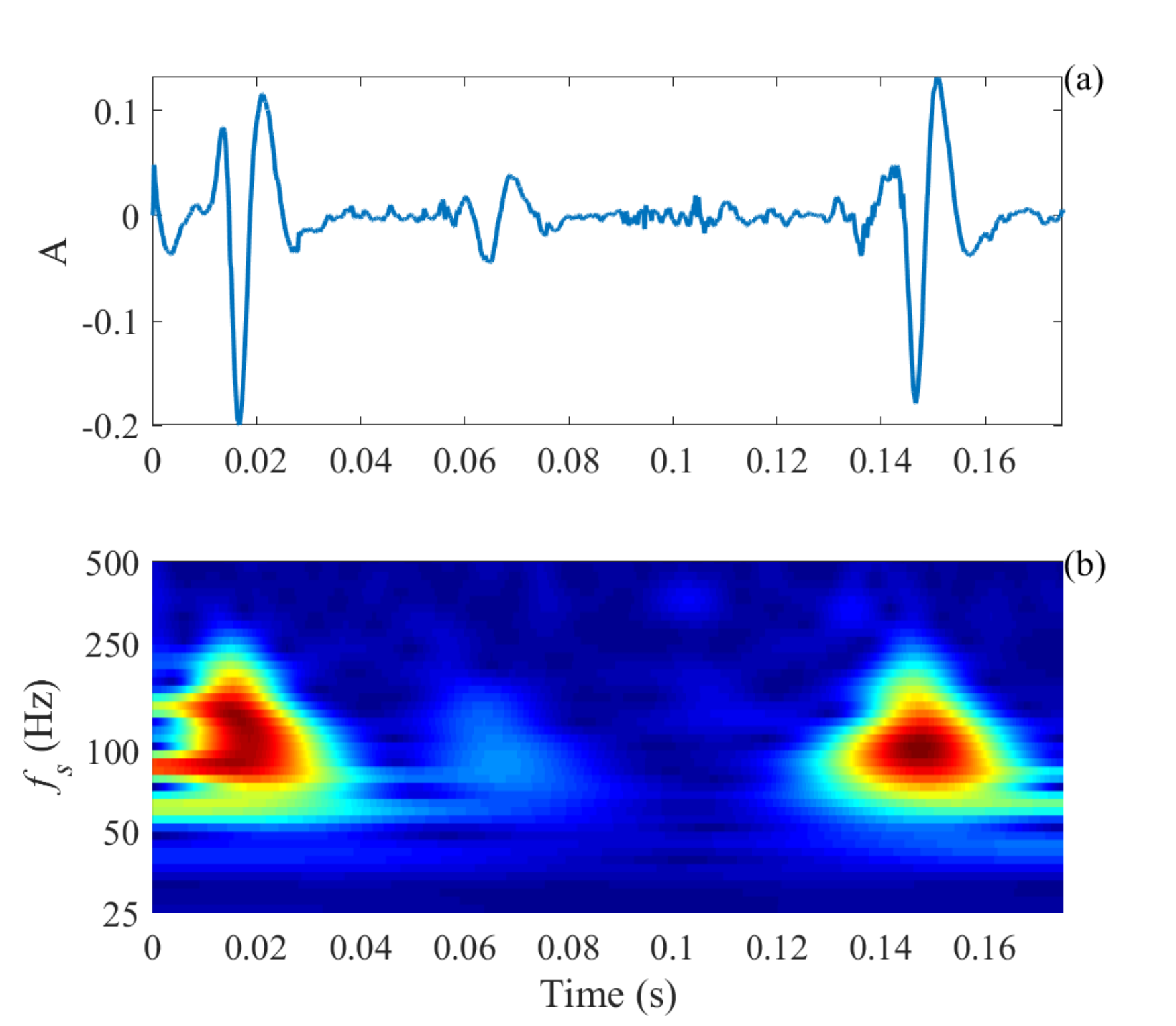}
\caption{(a) An example of an audio signal showing two beats ($r_d\approx1.8$ cm, $T_s=500^\circ$ C), (b) wavelet transform of the audio signal shown in (a).} 
\label{cwt_audio}
\end{figure}

\cref{expsp}(a) shows the experimental setup that was used to record the beats of a Leidenfrost drop. Experiments were performed by placing a drop of water on a concave surface (8cm radius of curvature) of aluminum. A heater from Thermo Scientific was used to keep the substrate at high temperature. The temperature of the substrate was obtained by using a FLUKE's thermocouple fixed to the substrate. A microphone (48 kHz sampling rate), connected to a regular smartphone, was held near the boundary of the drop to record the sound. A video camera (240fps) was fixed above the drop to record the videos. \cref{expsp}(b) shows some images of different geometrical patterns in the drop. As already mentioned, these are the modes of the drop which emit periodic beats. \cref{expsp}(c) shows a time series of an audio signal recorded from one of the experiments. This time series consists of two components: periodic spikes and interspike intervals. The spikes represent the beats emitted from the drop and the interspike interval part is the background noise. The duration of the interspike intervals for any audio signal corresponds to $f_b$. The background noise in the audio signal can have sources of various origins like electrical noise from the microphone, noise from the environment etc. The analysis of the audio signals has been done by filtering out the high frequency components of the background noise. Low frequency components of the noise could not be removed because the amplitude of the noise (in this frequency range) was comparable to the amplitude of the signal of interest and it was difficult to filter noise without altering the audio signal, and henceforth changing $f_s$. In \cref{expsp}(d) a time series representing the oscillations of the drop has been shown. A video of the drop was first recorded and the method of cross correlation \cite{Singla} was further employed on this video to obtain this time series. Finally, in \cref{expsp}(e), frequency spectra of time series shown in (c) and (d) have been shown; blue curve represents the spectrum of the audio signal and the orange curve represents the spectrum of the oscillations in the drop. It can be seen that the principle frequency of the oscillations in drop ($f_d$) is almost half the frequency of the beats ($f_b$). This is analogous to the result reported by Ma et al. \cite{Ma} where they showed that the frequency of the oscillations of a Leidenfrost star ($f_d$) is half the frequency of the vibrations in the vapor layer ($f_v$) underneath. The identical nature of $f_b$ and $f_v$ indicates that the vapor layer is responsible for the emission of beats (sound) from the drop, i.e. when vapors escape from the layer beneath the drop, they interact with the surrounding air to produce the beats. We emphasize that we refer to the periodic emission of sound as beats and not to the Leidenfrost stars (as reported by Ma et al.).

As already mentioned, the beats of the drop can be decomposed into two frequencies: $f_b$ and $f_s$. Frequency $f_b$ was obtained by employing Fourier transform technique on the audio signals recorded from the experiment. However, $f_b=2f_d$ and it has been reported earlier that $f_d$ is not a function of the drop radius but of capillary length of the fluid used \cite{Ma}. Therefore, $f_b$ that we obtained for different modes of oscillations did not vary with the drop radius and had similar values (results not shown) as reported in previous works.

\begin{figure}[ht!]
\includegraphics[width=9cm,height=5.5cm]{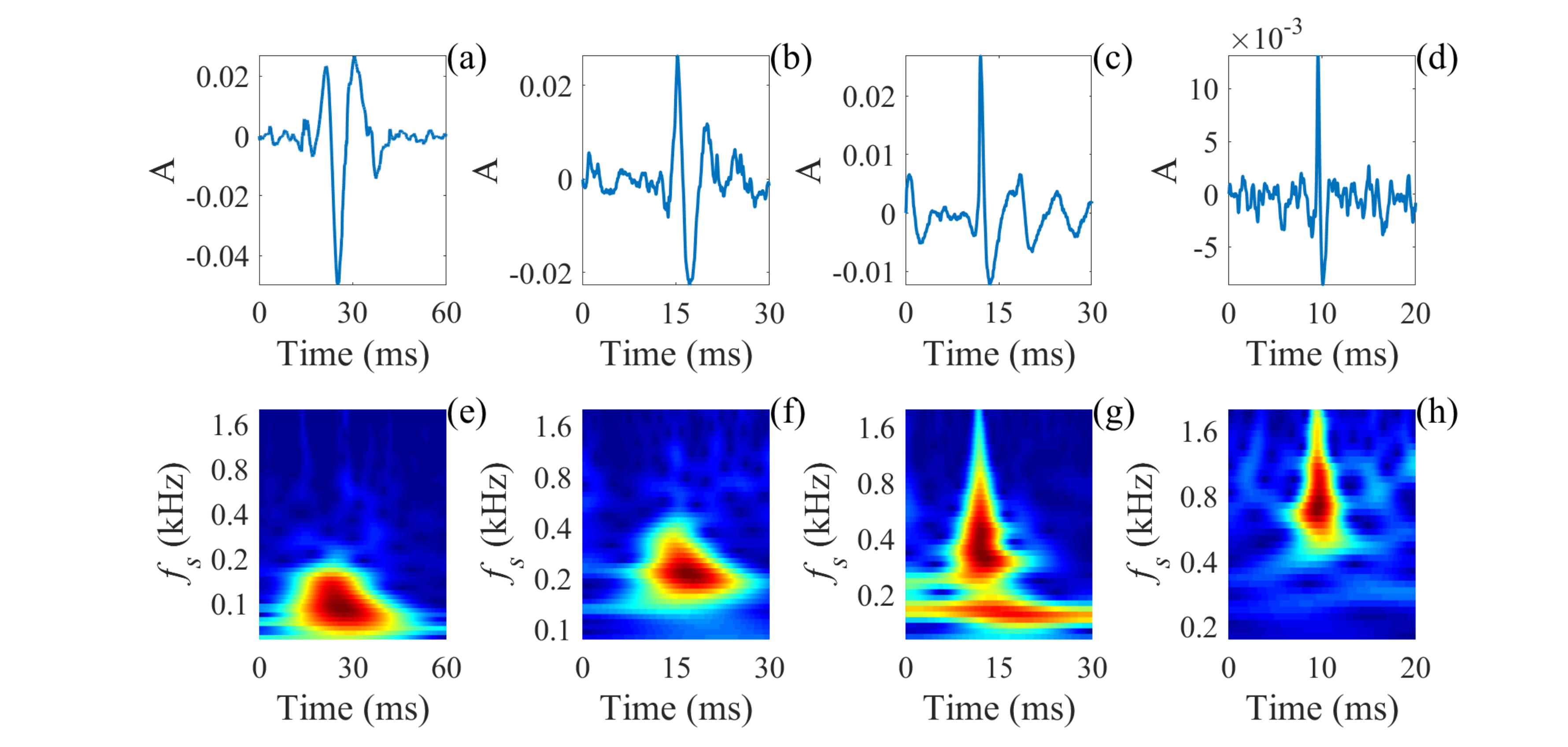}
\caption{Audio signals with corresponding wavelet transforms for four different drop sizes ($r_d$) and $T_s= 500^\circ$ C. $r_d$ for (a-d) were: (a) $1.69$ cm, (b) $1.42$ cm, (c) $0.61$ cm, (d) $0.33$ cm.} 
\label{cwt_audio1}
\end{figure}

To obtain the frequency of the sound ($f_s$), method of wavelet transforms was used (please refer to \cite{Supp} for more details on wavelet transforms). The technique of wavelet transforms allowed us to decompose the audio signals into frequency components at different times and a two dimensional frequency-time spectrum was obtained. In \cref{cwt_audio}(a), an audio signal consisting of two consecutive beats has been shown. The wavelet transform of this signal is shown in \cref{cwt_audio}(b). The information about frequencies present in the signal at any time moment can thus be obtained from these wavelet transforms; dark red color corresponding to dominant frequencies and blue color corresponding to insignificant frequencies of the signal. Therefore, the frequency of sound ($f_s$) in each of the beats can be obtained from these islands in the wavelet transform, which in the present case is $f_s\approx100$ Hz.

\cref{cwt_audio1}(a-d) show audio signals of a single beat for different drop sizes at $T_s=500^\circ$ C. The respective wavelet transforms of these signals have been shown in \cref{cwt_audio1}(e-h). These transforms show that as the size of the drop decreases, $f_s$ increases. We already know that the source of the beats is the vapor layer underneath the drop (\cref{expsp}). Therefore, sound emitted by Leidenfrost drop is identical to the acoustic modes of wind musical instruments where the fundamental frequency of the sound is given by:

\begin{equation}
f_s=\frac{v}{2l}
\label{eq1}
\end{equation}
here, $v$ is the velocity of sound and $l$ is the length of the instrument. Although, \cref{eq1} is commonly used for instruments being played by  human, the frequency of sound produced also depends on the pressure of the wind blown into the pipe \cite{Mallock}. Therefore, this relation requires a correction factor which, in the present case, incorporates the pressure of water vapor present beneath the drop. It has been reported that $f_s$ of wind instruments is proportional to the wind pressure \cite{Mallock}. For the sake of simplicity, assuming a linear dependence of the frequency of a wind instrument on wind pressure, \cref{eq1} transforms into:
\begin{equation}
f_s=\Big(\frac{p_v}{p_h}\Big)\frac{v}{2l}
\label{eq2}
\end{equation}
here, $p_v$ is the vapor pressure beneath the Leidenfrost drop and $p_h$ is the pressure of wind blown by human. In the present case, $v=691$ m/s \cite{Supp} at $T_s=500^\circ$ C, $l=r_d\approx0.61$ cm (\cref{cwt_audio1}(c)), and $p_v/p_h=1/100$ \cite{Chetta,Ma}. This gives $f_s\approx566$ Hz, which is roughly of the order of $f_s$ observed in \cref{cwt_audio1}(g). Moreover, it is also known that the thickness of the vapor layer beneath the drop is proportional to $r_d$ \cite{Burton}. Therefore, as the thickness of the vapor layer reduces in size, the vapors escape with higher velocities and interact differently with the surrounding air to produce sounds of higher frequencies. This is similar to the situation where a gas contained in a chamber escapes from a small hole and the frequency of the sound produced is inversely proportional to the size of the hole \cite{Anderson}. Another interesting fact that can be observed from \cref{cwt_audio1} is how the profile of audio signals changes as $r_d$ reduces. This behavior of the audio signals indicates different mechanism of beat emission in these cases; which is due to the fact that the morphology of the vapor layer for the drops with big $r_d$ could be quite distinct from that of small $r_d$.

\begin{figure}[ht!]
\includegraphics[width=8cm,height=3.5cm]{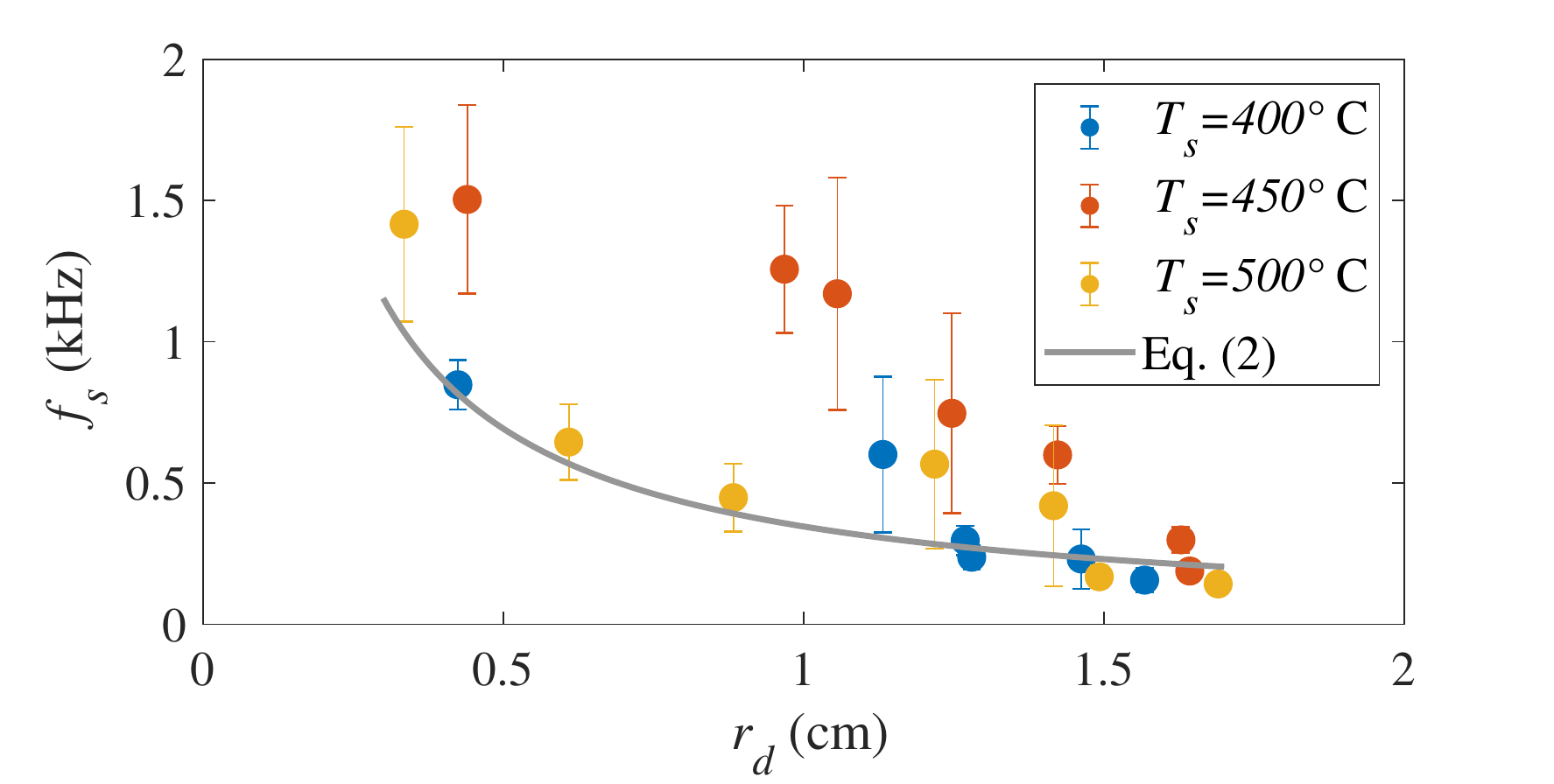}
\caption{Variation of $f_s$ as a function of $r_d$ and substrate temperatures. Parameters corresponding to the curve representing \cref{eq2} were: $T_s=500^\circ$C, $v=691$ m/s, $l=r_d$, and $p_v/p_h=1/100$.
} 
\label{err}
\end{figure}

The dependence of frequency of sound ($f_s$) on temperature of the substrate has been shown in \cref{err}. To explore this dependency, audio signals for several beats at different drop sizes ($r_d$) were obtained at three different temperatures. The $f_s$ of these audio signals along with corresponding errorbars were then plotted on \cref{err}. It can be observed that the $f_s$ does not show any definite relation with $T_s$. There could be two possible reasons for this behavior. First, it is possible that $f_s$ (similar to $f_d$) does not depend on $T_s$ at all. This, however, contradicts with \cref{eq2} which establishes relationship between $f_s$ and $T_s$. Second, it is possible that the change in $T_s$ in \cref{err} was not sufficient to significantly change $f_s$. This becomes evident when $f_s$ as estimated by \cref{eq2} is plotted on \cref{err} for different $T_s$. Gray curve shown on \cref{err} represents $f_s$ for $T_s=500^\circ$ C (identical curves were obtained for $T_s=450^\circ$ C and $T_s=400^\circ$ C). Moreover, as there exists a lot of background noise in the audio signals, clean audio signals could be extracted and analyzed only for few drop sizes. This, however, does not mean that beats were not detected at other drop sizes.

\begin{figure}[ht!]
\includegraphics[width=8cm,height=3cm]{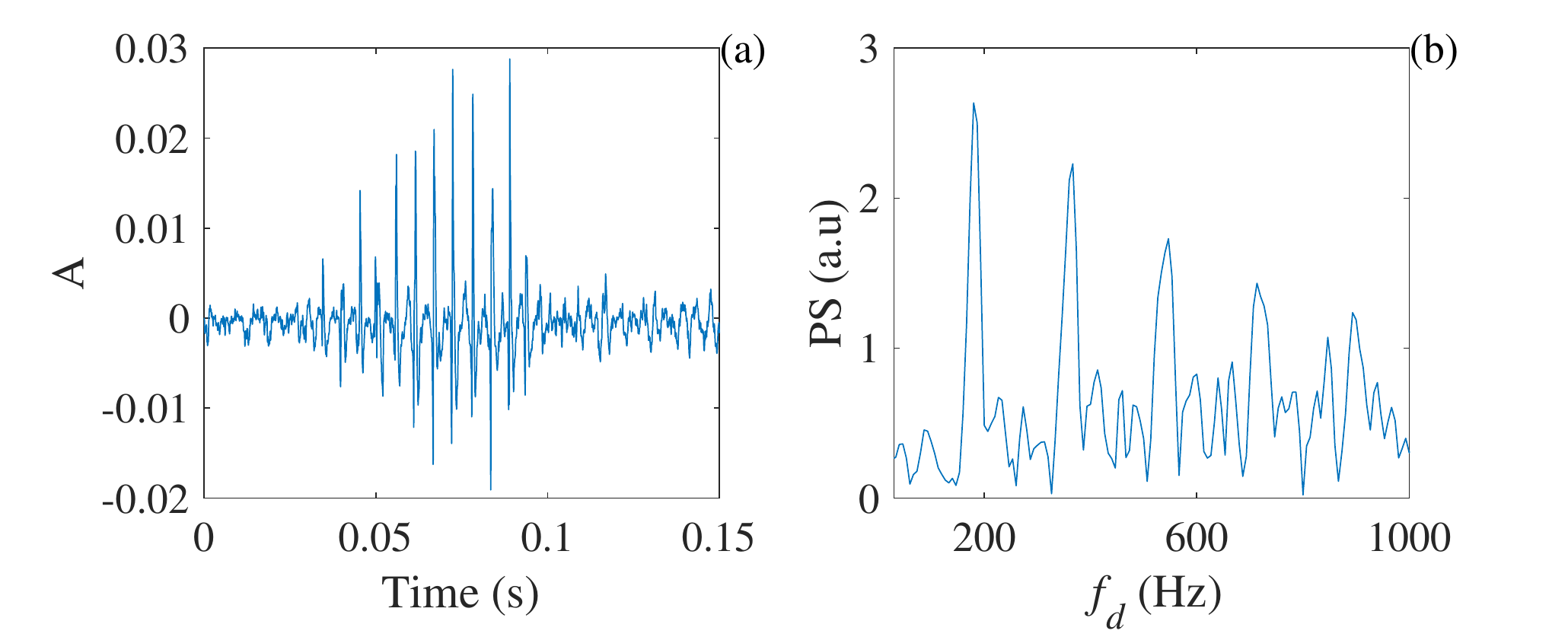}
\caption{Audio signal of beats when $r_d$ was very small such that it does not exhibit geometric oscillations ($r_d\approx1.5$ mm, $T_s=500^\circ$ C), (b) frequency spectrum of the audio signal shown in (a)}
\label{bounce}
\end{figure}

We now show results for beat emission from a small drop in which Leidenfrost stars cannot be observed. An example of an audio signal for this scenario along with its frequency spectrum has been shown in \cref{bounce} (a), and (b). The frequency spectrum shown in (b) has been obtained by manually removing the noise from the audio signal shown in (a). As can be seen in the frequency spectrum, the principle frequency of the beats in this case is around $180$ Hz. This signifies that the mechanism of beat emission in this case is distinct from the mechanism when the beats were emitted in the star configuration modes ($f_b\approx 23$ Hz). On closer inspection, it was observed that the drop in this case executed vertical bounces. A similar study was reported where  sound was emitted due to hydrogel spheres bouncing on a hot surface \cite{Waitukaitis,Waitukaitis1}. It was shown that the high frequency ``gap oscillations'' created by the impact of the sphere on the surface are responsible for this sound.  Waitukaitis et al. in \cite{Waitukaitis} explained gap oscillations as a cycle of three stage process. The first stage starts when the hydrogel ball makes initial contact with the hot surface. This causes the rapid vaporization of water and formation of a pocket beneath the drop where the vapors trap. The size of the pocket will keep increasing as long as the height of the pocket reaches to a threshold and at this point the drop lifts up due to vapor pressure. This marks the initiation of second stage in which vapors escape from the pocket causing recoiling of lower surface of the drop. With recoiling of the lower surface commences the third stage and it lasts as long as the drop makes another contact with the hot surface to restart first stage of the cycle.

Using the model of gap oscillations \cite{Waitukaitis}, we estimated gap oscillation frequencies for different drop sizes. It has been observed that this frequency decreases as the size of the drop decreases (please refer to \cite{Supp} for more details). With a diameter of $1.49$ cm of hydrogel spheres, Waitukaitis et al. reported a gap oscillation frequency of $f\approx 2.4$ kHz. In our case, the diameter of drop was around $3$ mm and $f\approx 180$ Hz. In our calculation of gap oscillations \cite{Supp}, the frequency corresponding to hydrogel sphere of diameter $3$ mm was $1.6$ kHz. The discrepancy of this result with the frequency of gap oscillation ($180$ Hz) could be due to the fact that the parameters used in the model were for hydrogel spheres (which although is 99\% water, behave like elastic solids).  To get the frequency of gap oscillations of a water drop, the appropriate parameters must be used in the model. Finally, for the sake of completeness, $f_s$ of the beats was also calculated for this case, and it had a value of $f_s\approx 1$ kHz (results not shown).

In conclusion, using basic experimental setup we showed the existence of periodic beats emerging from a Leidenfrost drop (please refer to \cite{Supp} for the audio recordings of the beats for different cases). In \cite{Brunet1} Philippe Brunet speculated the possible existence of sounds from Leidenfrost stars. However, Brunet mentioned if sounds with $f_b>50$ Hz can be generated from the drops. In our results, while the sound emission has been observed, $f_b$ remained in the order of $23$ Hz (which is the typical vapor layer frequency of the Leidenfrost stars). 

We used simple but powerful analytical techniques to qualitatively study the beats. These techniques allowed us to provide an insight about the process of beat emission, which otherwise could have required complicated experimental measures; typically used to study Leidenfrost drops. We showed that the vapors escaping from the layer beneath the drop are responsible for producing the beats and that the frequency of the sound increases as radius of drop decreases. Under these conditions the beats of the drop are equivalent to the acoustic modes produced by wind instruments. The analogy of beats with these acoustic modes was supported by our simple model represented by \cref{eq2}. A possible deviation of \cref{eq2} with the results of \cref{err} has also been mentioned. This indicates that a more sophisticated model should be developed to explain the beats of Leidenfrost drops. Finally, we showed that the gap oscillations of a bouncing water drop can result in the emission of periodic beats from the drop. 

\bibliography{biblio}
\end{document}